\documentclass{tPHM2e}

\usepackage{epstopdf}% To incorporate .eps illustrations using PDFLaTeX, etc.
\usepackage{subfigure}% Support for small, `sub' figures and tables

\theoremstyle{plain}

\theoremstyle{definition}

\theoremstyle{remark}

\begin{document}

%\jvol{00} \jnum{00} \jyear{2015} \jmonth{January}

%\articletype{SCES 2016 Conference Paper (Invited Talk)}

\title{Peculiar properties of Cr$_3$As$_3$-chain-based superconductors}

\author{
\name{Guang-Han Cao,\textsuperscript{a,b,}$^{\ast}$\thanks{$^\ast$Corresponding author. Email: ghcao@zju.edu.cn}
Jin-Ke Bao,\textsuperscript{a,b,}$^\dagger$\thanks{$^\dagger$Present Address: Materials Science Division, Argonne National Laboratory, Argonne, Illinois 60439, USA}
%Hao-Kun Zuo,\textsuperscript{b}
Zhang-Tu Tang,\textsuperscript{a,b}
Yi Liu,\textsuperscript{a,b}
Hao Jiang,\textsuperscript{a,b}$^\dagger$\thanks{$^\ddag$Present Address: School of Physics and Optoelectronics, Xiangtan University, Xiangtan 411105, China}
%Zeng-Wei Zhu,\textsuperscript{b}
%Yi Zhou,\textsuperscript{a,c}
%and Fu-Chun Zhang\textsuperscript{a,c}
}
\affil{\textsuperscript{a}Department of Physics, Zhejiang University, Hangzhou 310027, China;
%\textsuperscript{b}Wuhan National High Magnetic Field Center, School of Physics, Huazhong University of Science and Technology, Wuhan 430074, China;
\\
\textsuperscript{b}Collaborative Innovation Centre of Advanced Microstructures, Nanjing 210093, China}
\received{\today}
}

\maketitle

\begin{abstract}
$A_2$Cr$_3$As$_3$ ($A$ = K, Rb, Cs) are the unique Cr-based ambient-pressure superconductors to date discovered by serendipity in 2015. The new superconducting family are structurally characterized by quasi one-dimensional [(Cr$_3$As$_3$)$^{2-}$]$_{\infty}$ double-walled subnanotubes, which exhibit peculiar properties that mostly point to unconventional superconductivity. In this conference paper, we first describe how the superconductors were discovered. Then we overview the recent progresses on crystal structure, electronic structures, theoretical models, and various physical properties in $A_2$Cr$_3$As$_3$. Some new experimental results are included. Finally we conclude by addressing the related open questions in this emerging subfield of superconductivity.

\end{abstract}


\section{Introduction}

Unconventional superconductivity has been and remains one of the major topics in the area of condensed matter physics. According to the classification of known superconductors (SCs) by Hirsch, Maple and Marsiglio\cite{maple}, there are eleven classes of materials that can be categorized to be ``unconventional superconductors" (USCs) in which the superconductivity does not come from conventional electron-phonon interactions. These USCs include heavy-fermion compounds, organic charge-transfer compounds, cuprates, the ruthenate Sr$_2$RuO$_4$, U-based superconducting ferromagnets, cobalt oxyhydrates, iron-based SCs, etc.

Notably, the USCs frequently bear reduced dimensionality and strong electron correlations. While many of them contain quasi two-dimensional (Q2D) structural layers that generate superconductivity, it is very rare to find superconductivity in a material with quasi one-dimensional (Q1D) structure and, in particular, simultaneously with significant electron-electron interactions. Among the USCs listed in Ref.~\cite{maple}, only organic Bechgaard salts\cite{jerome} shows Q1D characteristic. Other Q1D superconductors include Li$_{0.9}$Mo$_{6}$O$_{17}$\cite{greenblatt} and Tl$_2$Mo$_6$Se$_6$\cite{armici}. Note that recent studies also suggest unconventional superconductivity in the former material\cite{denlinger,xxf,mercure,lebed13}.

In 2015, we reported bulk superconductivity at $T_{\mathrm{c}}$ = 6.1 K in K$_2$Cr$_3$As$_3$ whose crystal structure is Q1D\cite{K233}. This is also the first observation of superconductivity in Cr-based compounds at ambient pressure. Following this discovery, we found additional two superconducting members in the series, Rb$_2$Cr$_3$As$_3$ ($T_{\mathrm{c}}$ = 4.8 K)\cite{Rb233} and Cs$_2$Cr$_3$As$_3$ ($T_{\mathrm{c}}$ = 2.2 K)\cite{Cs233}. Since then, considerable research efforts have been made, which yields fruitful and important progresses in the subfield. In this conference paper we first let the readers know the route to the unexpected discovery. The non-superconducting ``cousins" $A$Cr$_3$As$_3$ ($A$ = K, Rb, Cs) found subsequently are introduced for comparison. Then we focus on the peculiar structural, electronic, superconducting and normal-state properties of $A_2$Cr$_3$As$_3$. Finally we summarize with some open questions to be addressed in the future.

\section{Discovery and crystal structure}

\subsection{Discovery of $A_2$Cr$_3$As$_3$ superconductors}
As is known, many 3$d$ transition-metal elements carry localized magnetic moments, which was earlier considered detrimental to superconductivity. The discovery of high-temperature superconductivity in cuprates\cite{bednorz}, in which Cu$^{2+}$ has a spin of 1/2, broke the previous research paradigm. Since then, much attention has been paid for 3$d$-element based compounds to find novel superconductors. Among the major discoveries along this line, the most prominent one is the birth of iron-based high-temperature superconductors\cite{hosono,cxh}, which has triggered tremendous research in recent years\cite{johnston,cxh2}. We were also stimulated to explore superconductivity in other 3$d$-element based compounds including MnAs-layer containing materials. Consequently, we observed an insulator-to-metal transition in hole doped La$_{1-x}$Sr$_x$MnAsO which hosts a large thermoelectric effect\cite{syl}. We also found an unusual ferromagnetic metallic state in heavily hole-doped Ba$_{1-x}$K$_x$Mn$_2$As$_2$ crystals\cite{bjk}. This novel state was later identified as an exotic half metallic state in which itinerant ferromagnetism coexists with the Mn-spin antiferromagnetism\cite{johnston2,johnston3}.

Among 3$d$ transition-metal compounds, no superconductivity has been observed in chromium- and manganese-based materials until the recent discovery of superconductivity in simple binary CrAs\cite{ljl,CrAs-jp} and MnP\cite{MnP} under high pressures. The superconductivity appears in the vicinity of magnetism, suggesting unconventional superconductivity in these compounds. It is thus of great interest to seek for ambient-pressure superconductivity in Cr- or Mn-based compounds. We tried to work on CrAs-layer based systems, which leads to the discovery of a few new metallic layered compounds including Sr$_2$Cr$_3$As$_2$O$_2$\cite{jh} and Ba$_2$Ti$_2$Cr$_2$As$_4$O\cite{ab}. Among CrAs-layer containing materials, the ``122"-type BaCr$_2$As$_2$\cite{Ba122Cr,singh} is the simplest. With the consideration of trivalence for Cr in superconducting CrAs (under pressures), therefore, we investigated the hole doping effect in Ba$_{1-x}$K$_x$Cr$_2$As$_2$, similar to our previous study in Ba$_{1-x}$K$_x$Mn$_2$As$_2$\cite{bjk}. We found that the K-for-Ba solubility limit is too low to tune the properties (say, to suppress the antiferromagnetic ordering, etc) of the parent material. We wondered whether the end member ``KCr$_2$As$_2$" can be synthesized (note that the powder X-ray diffraction database ICCD does not show the existence of KCr$_2$As$_2$).

\begin{figure}
\centering
\includegraphics[width=12cm]{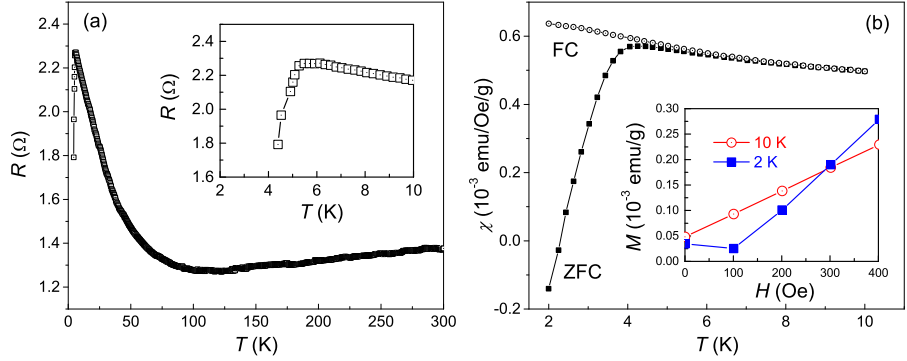}
\caption{Signature of superconductivity by resistance (a) and magnetic (b) measurements in ethanol-washed needle-like samples from the first grown batch of ``KCr$_2$As$_2$".}
\label{fig1}%\ref{fig:example1}
\end{figure}

The attempt to synthesize KCr$_2$As$_2$ eventually led to the discovery of the first Cr-based ambient-pressure superconductor K$_2$Cr$_3$As$_3$. We first tried to prepare KCr$_2$As$_2$ polycrystals via conventional solid-state reactions in a sealed quartz tube. However, the final product was mostly CrAs. Obviously, K was lost during the high-temperature reactions. Later, growing single crystals using CrAs flux also failed. Unexpectedly, needle-like crystals were seen when growing ``KCr$_2$As$_2$" crystals using KAs flux. To remove the flux attached (this procedure is actually unnecessary), the sample was ``cleaned" by being soaked in pure ethanol before performing physical-property measurements. Fig.~\ref{fig1}(a) show temperature dependence of resistance of the ``KCr$_2$As$_2$" sample. A resistance drop at about 5 K is seen, suggesting existence of trace of superconductivity. Fig.~\ref{fig1}(b) plots the dc magnetic measurement data for the sample, which indeed shows a magnetic susceptibility drop at about 4 K in the zero-field-cooling (ZFC) mode. The low-field magnetic susceptibility value becomes negative at 2 K, as confirmed by the $M-H$ curve shown in the inset. Since no superconductivity had been reported in K$-$As system, we believed that the superconductivity should come from a new superconducting phase that contains chromium.

\begin{figure}
\centering
\includegraphics[width=12cm]{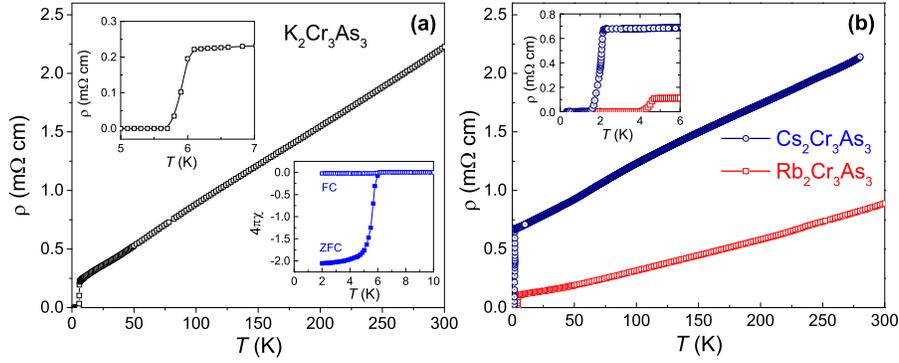}
\caption{(a) Superconductivity at 6.1 K in K$_{2}$Cr$_{3}$As$_{3}$ polycrystals, demonstrated by resistivity and dc magnetic susceptibility (the lower inset) measurements. (b) Temperature dependence of resistivity of Rb$_{2}$Cr$_{3}$As$_{3}$ and Cs$_{2}$Cr$_{3}$As$_{3}$ polycrystalline samples. The inset zooms in the superconducting transitions at 4.8 and 2.2 K, respectively. Data replotted from Refs.\cite{K233,Rb233,Cs233}}
\label{fig2}%\ref{fig:example1}
\end{figure}

To isolate and identify the new superconductive phase, we aimed at growing relatively large crystals. To avoid the loss of K, we employed sealed Ta tubes as the growth container. The need-like crystals obtained indeed show bulk superconductivity\cite{K233}. The chemical composition of the grown crystals was determined to be K:Cr:As = 2:3:3. Using this atomic ratio, we were able to synthesize a nearly single-phase ploycrystalline sample. As shown in Fig.~\ref{fig2}(a), the superconducting transition takes place at 6.1 K for the K$_{2}$Cr$_{3}$As$_{3}$ polycrystals. The diamagnetic signal (without demagnetization correction) in ZFC mode indicates bulk superconductivity. Note that the linear normal-state resistivity from 7 to 300 K is observed only for well-protected polycrystalline samples. Below we will show that the linear resistivity appears accidentally primarily due to the resistivity anisotropy.

Following the discovery of K$_{2}$Cr$_{3}$As$_{3}$ superconductor, we immediately attempted to make elemental replacements, in order to find more superconductors in this family, hopefully with higher $T_\mathrm{c}$. Consequently, we were able to find two sister superconductors, Rb$_{2}$Cr$_{3}$As$_{3}$ and Cs$_{2}$Cr$_{3}$As$_{3}$, which have lower $T_\mathrm{c}$ values of 4.8 K and 2.2 K, respectively. The trend in $T_\mathrm{c}$ implies that smaller cations at the $A$ site might yield higher $T_\mathrm{c}$. Unfortunately, synthesis of the target compounds such as Na$_{2}$Cr$_{3}$As$_{3}$ and Li$_{2}$Cr$_{3}$As$_{3}$ were unsuccessful. By the way, attempts to synthesize Mn-based analogues such as K$_{2}$Mn$_{3}$As$_{3}$ were unsuccessful either.

\subsection{Crystal structure and its relation to $T_\mathrm{c}$}

\begin{figure}
\centering
\includegraphics[width=8cm]{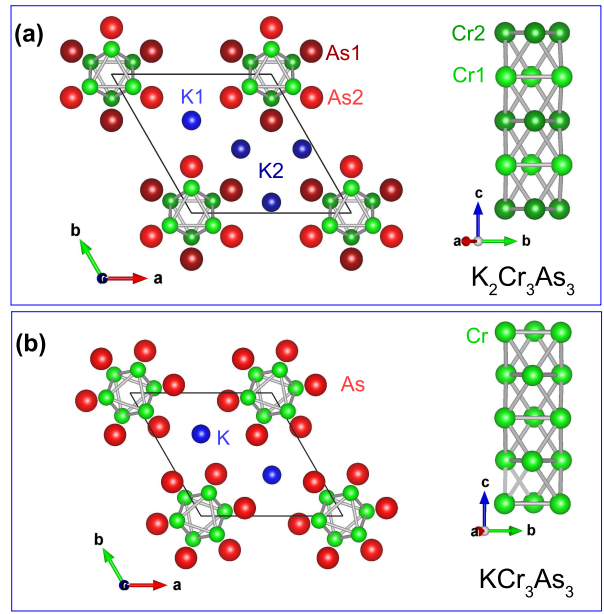}
\caption{Crystal structure of K$_{2}$Cr$_{3}$As$_{3}$ (a) and KCr$_{3}$As$_{3}$ (b) projected along the $c$ axis. The structure geometry of linear Cr$_{3}$ tube or chain (face-sharing Cr$_{6}$ octahedra) is highlighted at the right side, respectively.}
\label{fig3}%\ref{fig:example1}
\end{figure}

From crystal-chemical point of view, we figured out that Cr atoms should bond covalently with As, whereas As should bond ionically with K$^+$, separating electropositive Cr and K atoms. Furthermore, the crystal structure should be Q1D in accordance with the needle-like morphology. This speculation is verified by the crystal-structure determination using single-crystal X-ray diffractions\cite{K233}. As shown in Fig.~\ref{fig3}(a), the structure contains infinite [(Cr$_3$As$_3$)$^{2-}$]$_{\infty}$ linear chains or double-walled subnanotubes (DST) which are connected by K$^+$ cations. The [(Cr$_3$As$_3$)$^{2-}$]$_{\infty}$ DST is composed of inner Cr$_3$ twisted tubes (in green) and outer As$_3$ ones (in red), which are constructed by the face-sharing Cr$_6$ (or As$_6$) octahedra along the crystallographic $c$ direction. For the Cr sublattices that may carry magnetic moments, therefore, one expects strong geometrical magnetic frustrations.

The [(Cr$_3$As$_3$)$^{2-}$]$_{\infty}$ DSTs and the K$^+$ counterions form a hexagonal lattice with the space group of $P\overline{6}$$m$2. Every unit cell contains two formula units ($Z$ = 2), namely, the chemical formulae of one unit cell is formally K$_{4}$Cr$_{6}$As$_{6}$. Note that all the atoms occupy in the crystalline planes of $z$ = 0 and 0.5\cite{K233}. Two crystallographically different K sites, K1 (3$k$) and K2 (1$c$), locate at $z$ = 0.5 and 0, respectively. This atomic arrangement directly leads to the absence of inversion symmetry as well as the loss of six-fold rotation symmetry. Correspondingly, there exist two inequivalent As and Cr sites. In particular, the side length of Cr1 ($z$ = 0.5) triangles is 0.08(1) $\mathrm{\AA}$ shorter than that of the Cr2 triangles, implying possible difference in magnetic moments of Cr1 and Cr2\cite{hjp1}. Note that the Cr1$-$Cr2 distance is almost the same as that of Cr1$-$Cr1, reflecting strong chemical bonding along the $c$ axis.

\begin{figure}
\centering
\includegraphics[width=14cm]{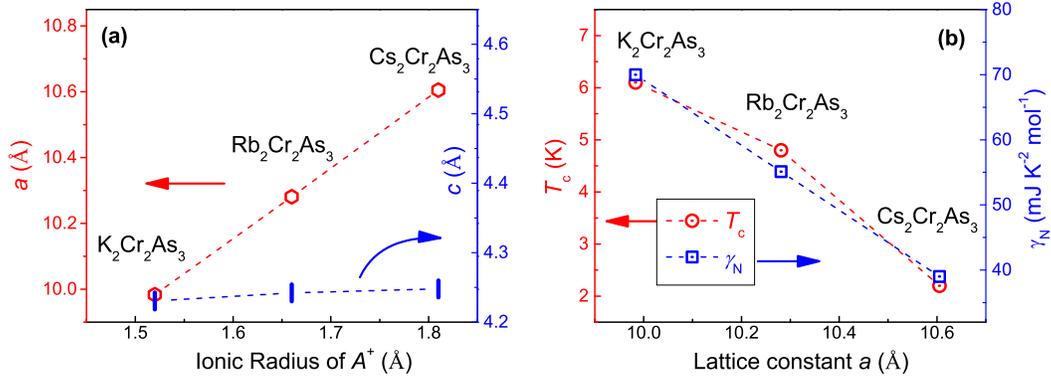}
\caption{(a) Lattice parameters as functions of the ionic radii of alkali-metal elements. (b) Superconducting transition temperature $T_\mathrm{c}$ and the electronic specific-heat coefficient versus the lattice parameter $a$ for  $A_{2}$Cr$_{3}$As$_{3}$ ($A$ = K, Rb, Cs).}
\label{fig4}%\ref{fig:example1}
\end{figure}

With the K$_{2}$Cr$_{3}$As$_{3}$ structure model, the crystal structures of Rb$_2$Cr$_3$As$_3$ and Cs$_2$Cr$_3$As$_3$ were solved by Rietveld refinements of the powder X-ray diffractions. Table~\ref{tab1} compares the structural parameters of $A$$_2$Cr$_3$As$_3$. With increasing the size of alkali-metal ions, the lattice constant $a$ (which also signifies the Cr$_3$As$_3$-interchain distance) increases remarkably, whereas the lattice constant $c$ axis increases slightly. This is clearly seen in Fig.~\ref{fig4}(a). Albeit of 6.0\% increase in $a$ from K$_2$Cr$_3$As$_3$ to Cs$_2$Cr$_3$As$_3$, the Cr$-$Cr bond distances basically remain unchanged. In contrast, the inter-chain Cr$-$Cr distance increases significantly because of the increase in $a$. Therefore, the structural trend tells us that the interchain coupling decreases with the lattice parameter $a$.

\begin{table}
\tbl{Crystallographic data of $A_2$Cr$_3$As$_3$ ($A$ = K, Rb, Cs). The atomic coordinates are as follows: As1 ($x$, $y$, 0); As2 ($x$, $y$, 0.5); Cr1 ($x$, $y$, 0.5); Cr2 ($x$, $y$, 0); $A$1 ($x$, $y$, 0.5); $A$2 (1/3, 2/3, 0). The space group is $P\overline{6}$$m$2.}
{\begin{tabular}[l]{@{}lcccccc}
\toprule
  Compounds & K$_2$Cr$_3$As$_3$\cite{K233}& & Rb$_2$Cr$_3$As$_3$\cite{Rb233} && Cs$_2$Cr$_3$As$_3$\cite{Cs233}& \\
\colrule
$a$ (\r{A})  & 9.9832(9)& &10.281(1)&&10.605(1)& \\
$c$ (\r{A})  & 4.2304(4)& &4.2421(3)&&4.2478(5)&\\
$V$ (\r{A}$^{3}$)    &365.13(6)& & 388.32(5)& & 413.73(9)&\\
\hline
Coordinates&$x$&$y$&  $x$&$y$ &  $x$&$y$\\
\hline
As1 (3$j$)&0.8339(2)&0.1661(2)& 0.8382(2)&0.1618(2) &  0.8399(3)&0.1601(3)\\
As2 (3$k$)&0.6649(4)&0.8324(4)&  0.6727(2) &0.8364(2)&0.6708(3)&0.8354(3) \\
Cr1 (3$k$)&0.9127(3)&0.0873(3) & 0.9140(2) & 0.0860(2)&0.9143(5)&0.0857(5)\\
Cr2 (3$j$)&0.8203(6)&0.9102(6) &  0.8333(3) & 0.9167(3)&0.8375(4)&0.9187(4)\\
$A$1 (3$k$)&0.5387(4)&0.0775(4)&   0.5372(1) &0.0744(1)&0.5328(2)&0.0655(2) \\\hline
Interatomic Distances && & & & &\\\hline
Cr1$-$Cr1 (\r{A}) & 2.614(9) & &   2.654(4) & &2.718(1) &\\
Cr1$-$Cr2 (\r{A}) & 2.612(2) &  &  2.603(2) & &2.621(2)& \\
Cr2$-$Cr2 (\r{A}) & 2.690(10) &   & 2.570(3) & & 2.585(4)&\\
Inter-chain Cr1$-$Cr1 (\r{A}) & 7.369(5) &   & 7.627(3) & &7.878(8) &\\
\botrule
\end{tabular}}
\label{tab1}
\end{table}

There exists inconsistency for the resulted Cr$-$Cr bond distances, which shows that $d_{\mathrm{Cr1-Cr1}}$ values of Rb$_2$Cr$_3$As$_3$ and Cs$_2$Cr$_3$As$_3$ are abnormally larger than $d_{\mathrm{Cr2-Cr2}}$ (see Table~\ref{tab1}), in contrast to the case in K$_2$Cr$_3$As$_3$. We note that such inconsistency is absent in the crystallographic data of Rb$_2$Cr$_3$As$_3$ obtained from the single-crystal X-ray diffractions\cite{jp}. Since there are three $A^+$ ions at the $z$=0.5 plane, the Cr1 triangles at $z$=0.5 plane should be compressed, which reasonably leads to a shorter $d_{\mathrm{Cr1-Cr1}}$ value. Hence there could be some systematic deviations in the Rietveld analyses of the powder X-ray diffraction data for Rb$_2$Cr$_3$As$_3$ and Cs$_2$Cr$_3$As$_3$ (note that the samples are very air sensitive, which may influence collections for the high-quality data).

Fig.~\ref{fig4}(b) shows the relation between $T_\mathrm{c}$ and the inter-chain distance (or the lattice parameter $a$) in $A_{2}$Cr$_{3}$As$_{3}$. $T_\mathrm{c}$ decreases monotonically with increasing $a$, suggesting roles of inter-chain coupling in determining $T_\mathrm{c}$. This trend seems to suggest that $T_\mathrm{c}$ can be increased by applying pressures [see Fig.~\ref{fig5}(c) below]. However, the high-pressure results show that $T_\mathrm{c}$ always \emph{decreases} with pressure in both K$_2$Cr$_3$As$_3$ and Rb$_2$Cr$_3$As$_3$\cite{canfield1,sll1}. The updated systematic study by employing structural determinations under high pressures concludes that the superconductivity has a strong positive correlation with the degree of non-centrosymmetry\cite{sll2}. Indeed, the absence of superconductivity in centro-symmetric $A$Cr$_3$As$_3$ (see below) seems to support this point of view.

\subsection{Absence of superconductivity in $A$Cr$_3$As$_3$}

The $A$$_2$Cr$_3$As$_3$ compounds are not stable, and they easily deteriorate chemically at ambient condition. By using a topotactic soft-chemical route that keeps the Cr$_3$As$_3$ chains undestroyed, we obtained the corresponding ``cousin" compounds $A$Cr$_3$As$_3$\cite{bjk2,tzt2}, which are stable in air. The resulted materials lose two $A$ ions in a unit cell, as shown in Fig.~\ref{fig2}(b). As a result, the lattice parameters $a$ (= 9.091 \AA) and $c$ (= 4.181 \AA) of KCr$_3$As$_3$ decrease by 9\% and 1\%, respectively, compared with those of K$_2$Cr$_3$As$_3$. More importantly, there is only one site for $A$ ions, which changes the point group from $D_{3h}$ to $D_{6h}$, and the space group from $P\overline{6}$$m$2 (No. 187) to $P6_{3}/m$ (No. 176). Another consequence is that there is also only one Cr (or As) site.

All the $A$Cr$_3$As$_3$ compounds do not superconduct, and they exhibit a cluster-spin-glass behavior below $\sim$5 K (close to the $T_\mathrm{c}$ values in $A_2$Cr$_3$As$_3$)\cite{bjk2,tzt2}. The spin-glass-like state in $A$Cr$_3$As$_3$ seems to be associated with the geometrical frustrations in the Cr$_3$ twisted tubes. At high
temperatures, $A$Cr$_3$As$_3$ shows a Curie-Weiss behavior, indicating existence of local moments. Owing to the close relation between $A$Cr$_3$As$_3$ and $A_2$Cr$_3$As$_3$, the local-moment magnetism in $A$Cr$_3$As$_3$ implies that Cr spins are relevant to the generation of superconductivity. In this respect, the absence of superconductivity in $A$Cr$_3$As$_3$ is as puzzling as the appearance of superconductivity in $A_2$Cr$_3$As$_3$. Here we summarize several possible explanations. One emphasizes the importance of unequal Cr sites\cite{hjp1,hjp2}, which is supported by the high-pressure study\cite{sll2}. However, there is an obvious difference in electron filling level. The apparent valence of Cr in 233-type system is +2.33 (assuming K$^+$ and As$^{3-}$), which means that the number of 3$d$ electrons of the chromium is 3.67. While the apparent Cr valence for the 133-type series is +2.67, namely, the number of Cr 3$d$ electrons is 3.33. This difference in electron filling seems to be crucial in determining the different ground states. First-principles calculations on KCr$_3$As$_3$ show that the ground state is an interlayer AFM order\cite{cc2}, which is in sharp contrast with the result for K$_2$Cr$_3$As$_3$\cite{cc1,hjp1}. In the presence of the AFM order and with the lower electron filling, the Fermi surface (FS) of KCr$_3$As$_3$ only involves three Q1D sheets. In this perspective, the absence of superconductivity could be related to the severely reduced dimensionality.

The above $A$Cr$_3$As$_3$ materials are reminiscence of the molybdenum cluster compounds $M_2$Mo$_6$$Ch_6$ or $M$Mo$_3$$Ch_3$ ($M$ = Na, K, Rb, Cs, In, TI; $Ch$ = S, Se, Te) \cite{potel,honle} and an iron telluride TlFe$_3$Te$_3$\cite{klepp}. In fact, they all share the identical crystal-structure type. However, the number of Mo 4$d$ electrons in $M$Mo$_3$$Ch_3$ is 4.33, one more electron than that in $A$Cr$_3$As$_3$. The differences in electron filling and electron correlations make the electronic structure different fundamentally\cite{petrovic,alemany}. Note that in the Mo-based chalcogenides, most show semiconducting/insulating behavior. Only TlMo$_3$Se$_3$ and InMo$_3$Se$_3$ exhibit superconductivity\cite{armici}, which are considered to be conventional\cite{petrovic}, though being strongly anisotropic. For the unique 133-type Fe-based telluride TlFe$_3$Te$_3$, a ferromagnetic ordering is formed below 220 K with the Fe spins parallel to the $c$ axis\cite{uhl,bronger}. Note that both the 3$d$-element based materials $A$Cr$_3$As$_3$ and TlFe$_3$Te$_3$ host magnetism, consistent with the significant electron correlations presented commonly in 3$d$-element systems.

\section{Electronic structure calculations and theoretical analysis}

First-principles electronic-structure calculations based on density functional theory (DFT) have become more and more important in understanding new materials. The discovery of superconductivity in $A_2$Cr$_3$As$_3$ have attracted several related calculations\cite{cc1,hjp1,alemany,subedi}, which not only depict the basic electronic structures, but also establish a basis for theoretical models\cite{hjp2,zy,djh,hjp3}. Below we briefly describe the progress in the theoretical and calculation aspects.

\subsection{On the magnetic ground state}
\begin{figure}
\centering
\includegraphics[width=10cm]{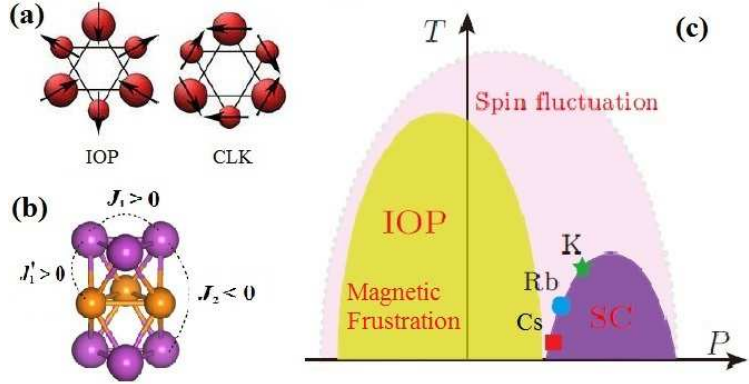}
\caption{(a) Non-collinear in-out co-planar (IOP) and chiral-like (CLK) magnetic configurations as possible ground states by DFT calculations. (b) Relevant exchange coupling parameters in which ferromagnetic $J_2$ is extracted. (c) Conjectured phase diagram for $A_{2}$Cr$_{3}$As$_{3}$ ($A$ = K, Rb, Cs). Adapted from Ref.\cite{hjp1}}
\label{fig5}%\ref{fig:example1}
\end{figure}

By calculating the total energies for different magnetic states, the possible ground state may be given. Considering the geometrical magnetic frustration, Wu et al.\cite{hjp1} propose a novel in-out co-planar (IOP) magnetic ordering [see the left of Fig.~\ref{fig5}(a)], apart from other possible collinear magnetically ordered states. They found that the IOP state is exclusively the ground state in K$_2$Cr$_3$As$_3$ and Rb$_2$Cr$_3$As$_3$. From the energy differences of various magnetic-order states, they were able to extract the magnetic exchange parameters within the Heisenberg model. The result shows that the next-nearest (NN) exchange couplings, intra-plane $J_1$ and inter-plane $J_1'$, are strongly antiferromagnetic (AFM), while the next NN exchange inter-plane coupling $J_2$ is ferromagnetic (FM) ($J_2<$0), as shown in Fig.~\ref{fig5}(b). Assuming a dominant intra-band pairing, the effective magnetic fluctuations within each Cr sublattice along the $c$ direction should be FM, which favors a spin-triplet Cooper pairing\cite{hjp1}. These authors further conjecture that superconductivity of $A_{2}$Cr$_{3}$As$_{3}$ with different $T_\mathrm{c}$ locates in the vicinity of the IOP phase [Fig.~\ref{fig5}(c)], and some sort of uniaxial pressure might push the $T_\mathrm{c}$.

Jiang et al.\cite{cc1} made a more systematic calculation for different magnetic states using both experimental and optimized structures, without and with the spin-obit coupling, and without and with plus $U$. The possible magnetic ordering includes an additional noncollinear chiral-like (CLK) state [see the right of Fig.~\ref{fig5}(a)]. What they found is that the IOP and CL states are degenerate, which explains the paramagnetic behavior observed in experiment. Meanwhile, the calculated bare electron susceptibility shows an extremely strong peaks at $\Gamma$ point for the imaginary part, suggesting strong FM spin fluctuations. The latter is supported by the NMR result on Rb$_2$Cr$_3$As$_3$ (see below).

\subsection{Band structure and Fermi surface}
\begin{figure}
\centering
\includegraphics[width=14cm]{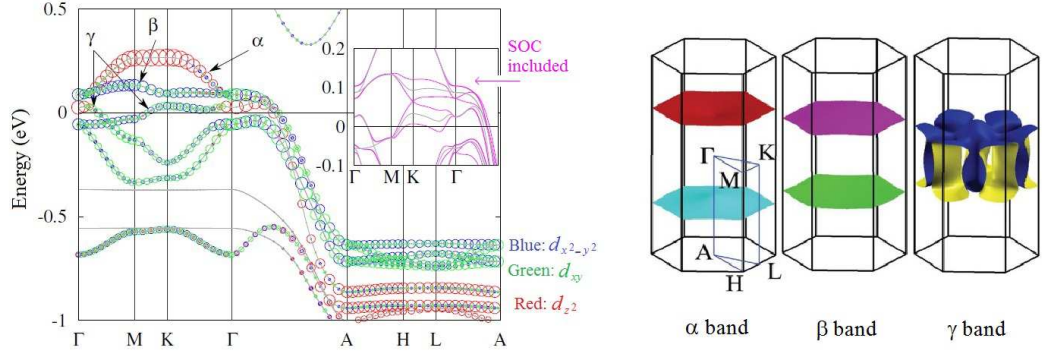}
\caption{Energy band structure of nonmagnetic K$_{2}$Cr$_{3}$As$_{3}$. The inset shows the spin-orbit-coupling (SOC) effect by comparing relativistic and non-relativistic caluculation results. Left panel shows the three Fermi surface sheets in $\alpha$, $\beta$ and $\gamma$ bands (non-relativistic results). Adapted from Ref.\cite{cc1}}
\label{fig6-FS}%\ref{fig:example1}
\end{figure}

The band-structure calculations for the prototype K$_2$Cr$_3$As$_3$ from different groups\cite{cc1,hjp1,alemany} basically give consistent results. Fig.~\ref{fig6-FS}(a) shows the band structure and the Fermi surface (FS) for nonmagnetic K$_2$Cr$_3$As$_3$. First of all, the electronic states near the Fermi level ($E_\mathrm{F}$) are almost exclusively dominated by three Cr-3$d$ orbitals ($d_{z^{2}}$, $d_{xy}$ and $d_{x^{2}-y^{2}}$). Secondly, the valence band width is only 0.6 eV at the $k_z$ = 0 plane, and it would be even more narrower when the electron-mass renormalization is considered. The narrowness of the valence bands implies significance of electron correlations in the Cr-based materials. Thirdly, there are three bands crossing $E_\mathrm{F}$, denoted as $\alpha$, $\beta$ and $\gamma$, respectively. $\alpha$ and $\beta$ bands only cut the $\Gamma-\mathrm{A}$ line, forming Q1D FSs (the FS sheets are almost planar, perpendicular to $k_z$), as shown in Fig.~\ref{fig6-FS}(b). On the other hand, $\gamma$ band not only cuts the $\Gamma-\mathrm{A}$ line, but also crosses the M$-$K line, which makes a 3D-like FS. In fact, the 3D FS can be regarded as a warping of two Q1D FS slices\cite{alemany}.

The total density of states (DOS) at $E_\mathrm{F}$ is calculated to be $N(E_{\mathrm{F}})$ = 8.58\cite{cc1} (8.76\cite{hjp1} and 6.27\cite{alemany}) eV$^{-1}$$\cdot$ fu$^{-1}$, corresponding to an electronic specific-heat coefficient of 20 (15) mJ K$^{-2}$ mol$^{-1}$. The experimental specific-heat value (70 mJ K$^{-2}$ mol$^{-1}$) is 3.5-4.5 times larger, indicating remarkable electron-mass renormalization. Here we also note that the 3D $\gamma$ band contributes 75\% of the DOS ($\alpha$ and $\beta$ bands contribute 8\% and 17\% respectively), which implies that this band might be dominantly responsible for superconductivity. Wu at al.\cite{hjp1} also find that Cr1 ions
have more $d$ electrons than Cr2 ions, which may lead to different magnetic moments at Cr1 and
Cr2 sites.

Fig.~\ref{fig4}(b) shows that there is a relation between $T_\mathrm{c}$ and the Sommerfeld coefficient $\gamma_\mathrm{N}$. However, the calculated DOS of $A_{2}$Cr$_{3}$As$_{3}$ do not decrease with increasing the ionic radius of $A^+$. Wu et al.\cite{hjp1} shows a DOS of 9.13 eV$^{-1}$$\cdot$ fu$^{-1}$ for Rb$_{2}$Cr$_{3}$As$_{3}$. Alemany and Canadell\cite{alemany} notice that, whereas $T_\mathrm{c}$ is reduced by 42\% from Rb$_{2}$Cr$_{3}$As$_{3}$ to Cs$_{2}$Cr$_{3}$As$_{3}$, the $N(E_{\mathrm{F}})$ remains practically constant. This seems to suggest that the electron-mass renormalization factor is obviously different for $A$ = K, Rb and Cs. Here we note that the detailed electronic structures for each $A_{2}$Cr$_{3}$As$_{3}$ differ remarkably, especially for the $\gamma$ band\cite{hjp1,alemany}. An additional Q1D FS is reported for Rb$_{2}$Cr$_{3}$As$_{3}$\cite{hjp1}.

\subsection{Theoretical model and analysis}

The DFT-based first-principles calculations supply a basis for constructing an effective theoretical model\cite{zy,hjp2,hjp4,djh,kim}. By solving the ``minimum" model and its extensions using suitable approximations, one may obtain information about the superconducting pairing mechanism especially for the pairing symmetry.

Zhou, Cao and Zhang\cite{zy} propose a minimal model with three molecular orbitals in each unit cell of K$_{2}$Cr$_{3}$As$_{3}$. The model well reproduces the FS and low-energy band structures from the first-principles calculations. Considering intraband pairings ($\mathbf{k}$,$-\mathbf{k}$), they give ten possible single-band superconducting gap functions within the lattice point group of $D_{3h}$, five of which are spin singlet and, the other five are spin triplet. The effective pairing interactions are found to be always most attractive in the spin-triplet channels. The pairing strength as functions of Hubbard $U$ and Hund's coupling $J_\mathrm{H}$ suggests two pairing candidates. At small $U$ and moderate $J_\mathrm{H}$, the pairing arises from 3D $\gamma$ band and has a spatial symmetry of $f_{y(3x^{2}-y^{2})}$. For a large $U$, however, a $p_z$ wave is stabilized within the Q1D band. According to their judgement, the former case is more likely for the Cr-based superconductors.

Wu et al.\cite{hjp2} present a similar tight-binding effective models based on the $d_{z^{2}}$, $d_{xy}$ and $d_{x^{2}-y^{2}}$ orbitals of the Cr2 sublattice, which also captures the band structures near $E_\mathrm{F}$ well. Their systematic and thorough calculations and analysis show that, in both weak and strong coupling limits, the triplet $p_z$-wave pairing is the leading pairing symmetry for physically realistic parameters, although they also find that $f_{y(3x^{2}-y^{2})}$-wave state can be stabilized for a large $J_\mathrm{H}/U$. Consistent with their previous speculation\cite{hjp1}, the triplet pairing is driven by the ferromagnetic fluctuations within the Cr2 sublattice. The conclusion of $p_z$-wave pairing is found to be robust when the model is extended to six bands which involve both Cr sublattices\cite{hjp5}. Experimental consequences of $p_z$-wave spin-triplet superconductivity are also given by the same authors\cite{hjp4}, which shows no obvious conflicts with the experimental observations to date.

Considering the central role of Q1D [(Cr$_3$As$_3$)$^{2-}$]$_{\infty}$ DSTs, Dai and co-workers\cite{djh} model the $A_{2}$Cr$_{3}$As$_{3}$ system as a twisted Hubbard tube. They exactly solve the molecular-orbital bands emerging from the quasi-degenerate atomic orbitals. With an effective three-band Hamiltonian for the low-energy region, the resulting three-channel Luttinger liquid shows various instabilities including two kinds of spin-triplet superconductivity due to gapless spin excitations.

The above theoretical works emphasize the important roles of electron-electron interactions and reduced dimensionality, which basically give consistent results that point to unconventional non-phonon-mediated superconductivity. However, there have been theoretical efforts to account for the superconductivity in terms of conventional electron-phonon interactions. Subedi\cite{subedi} studied the lattice dynamics and electron-phonon coupling in K$_{2}$Cr$_{3}$As$_{3}$ using DFT-based calculations. The result shows that the total electron-phonon coupling is as large as $\lambda_{\mathrm{ep}} = 3.0$, which readily explains the experimentally observed large mass renormalization. Wachtel and Kim\cite{kim} simplify the system into a two-band model. One of the bands is 1D, which couples with the other band of 3D. They find that the $2k_\mathrm{F}$ density fluctuations in the 1D band induces attractive interactions between the 3D electrons, making the system superconducting. As a result, a $d_{z^{2}}$-like gap function is obtained when strong enough local repulsion is included, which explains the experimental observations that point to nodal lines in the gap function.

\section{Normal-state and superconducting properties}\label{sec:4}

As is known, sample's quality is often crucial for acquiring intrinsic physical properties of a material. As for $A_{2}$Cr$_{3}$As$_{3}$, this aspect is especially important because of their extreme sensitivity to ambient conditions. As stated above, $A_{2}$Cr$_{3}$As$_{3}$ partially degrades into non-superconducting $A$Cr$_{3}$As$_{3}$ in the case of slow reaction with water in air. Therefore, cautions should be taken as far as possible to avoid the sample's deterioration during the handling in the physical-property measurements.

\subsection{Resistivity}
There has been an apparent discrepancy for the $\rho(T)$ behavior between polycrystals and single crystals of K$_{2}$Cr$_{3}$As$_{3}$. As shown in Fig.~\ref{fig2}(a), the $\rho(T)$ data of K$_{2}$Cr$_{3}$As$_{3}$ \emph{polycrystals} exhibit a strikingly linear behavior in a broad temperature range from 7 to 300 K. Similar linear $\rho(T)$ behaviors are also seen for Rb$_{2}$Cr$_{3}$As$_{3}$ and Cs$_{2}$Cr$_{3}$As$_{3}$ polycrystals, albeit in a narrow temperature range of $T_{\mathrm{c}}<T<$40 K\cite{Rb233,Cs233}. Canfield and co-workers\cite{canfield1} first reported the $\rho^{\parallel}(T)$ (with electric current parallel to the $c$ axis) data of K$_{2}$Cr$_{3}$As$_{3}$ crystals, which shows a power-law relation, $\rho(T)=\rho_{0}+AT^\alpha$, with $\alpha$= 3.0$\pm$0.2 for 10 K$<T<$40 K. We recently found that $\alpha$ decreases with impurity scattering\cite{ly}. For high-quality crystals with low residual resistivity $\rho_{0}$, the $\alpha$ value is indeed around 3.0. Then, what causes the linearity in the polycrystalline samples?

\begin{figure}
\centering
\includegraphics[width=12cm]{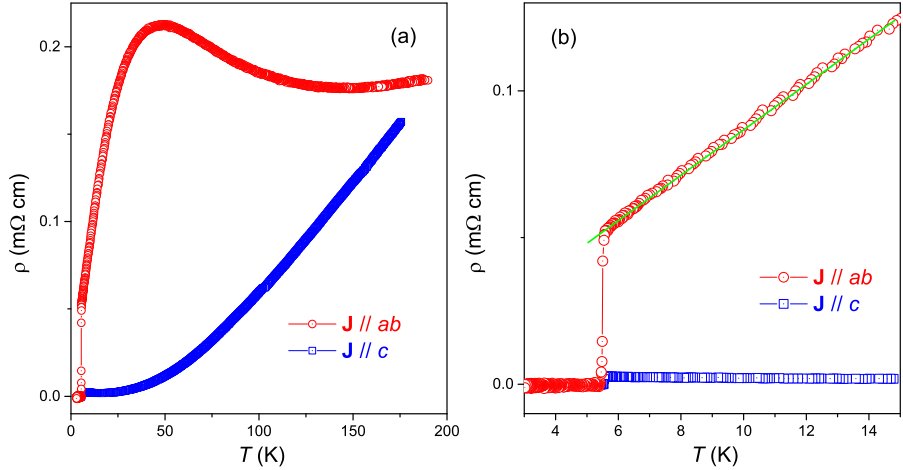}
\caption{ Temperature dependence of anisotropic resistivity (electric current flows along the $c$ direction or within the $ab$ plane) of the K$_{2}$Cr$_{3}$As$_{3}$ crystal. (a) shows a wide temperature range. (b) zooms in the low-temperature range below 15 K.}
\label{fig7}%\ref{fig:example1}
\end{figure}

To clarify the above discrepancy, it is very crucial to measure the resistivity with current flowing perpendicular to the $c$ axis (or within the $ab$ plane), signalled as $\rho^{\perp}(T)$. Fig.~\ref{fig7} shows both $\rho^{\perp}(T)$ and $\rho^{\parallel}(T)$ data for the same K$_{2}$Cr$_{3}$As$_{3}$ crystal. While the $\rho^{\parallel}(T)$ data almost show the same feature as the previous reports\cite{canfield1,ly} (although the $T_\mathrm{c}$ value is somewhat lowered possibly due to sample's partial degradation during the electrode preparation), the $\rho^{\perp}(T)$ data exhibit a very different behavior. There is an upturn below $\sim$150 K until it peaks at $\sim$50 K, below which it decreases rapidly till the superconducting transition. The high-temperature resistivity values are comparable for the two electric-current directions within the measurement uncertainty. Nevertheless, the resistivity anisotropy, $\gamma_{\rho}=\rho^{\perp}(T)/\rho^{\parallel}(T)$, achieves 120 and 10 at 50 K and 6 K, respectively, which manifests the Q1D nature for the material. Note that the $\rho^{\perp}(T)$ data indeed show a quasi-linear behavior below $\sim$30 K, which naturally explains the linearity in polycrystals. The linearity $\rho^{\perp}(T)$ just above $T_\mathrm{c}$ calls for theoretical explanations.

The semiconducting-like behavior in the high-$T$ range suggests incoherent transport within the $ab$ plane. If this is the case, the resistivity peak at $\sim$50 K corresponds to a dimension crossover from 1D to 3D with decreasing temperature. Therefore, one may speculate that in the high-$T$ region the system indeed shows a Luttinger liquid behavior, as is studied theoretically\cite{djh} and, suggested by the NQR measurement\cite{imai1}. Further investigations of the anisotropic transport properties are expected to clarify the nature of the normal state.

\subsection{Upper critical fields}
The upper critical field ($H_{\mathrm{c2}}$) of a type-II superconductor may give important hints to the superconducting pairing mechanism\cite{zhangwei}. In our original papers\cite{K233,Rb233,Cs233} we report the preliminary result of $H_{\mathrm{c2}}(T)$ for $A_{2}$Cr$_{3}$As$_{3}$ polycrystals. The $H_{\mathrm{c2}}(T)$ data actually represent some averages of $H_{\mathrm{c2}}^{\parallel}(T)$ (with $\mathbf{H}\parallel c$) and $H_{\mathrm{c2}}^{\perp}(T)$ (with $\mathbf{H}\perp c$). The initial slope of $H_{\mathrm{c2}}(T)$, $\mu_{0}$(d$H_{\mathrm{c2}}$/d$T)|$$_{T_{\mathrm{c}}}$, is as high as $-$7.43 T/K for K$_{2}$Cr$_{3}$As$_{3}$\cite{K233}, corresponding to a large orbitally limited $\mu_{0}H_{\mathrm{c2}}^{\mathrm{orb}}(0)$ of $\sim$32 T as estimated by Werthammer-Helfand-Hohenberg model\cite{WHH}. The latter is about three times larger than the Pauli-paramagnetic limit $\mu_{0}H_\mathrm{P}$ = 1.84 $T_\mathrm{c}$ = 11 T\cite{clogston,chandrasekhar}. Thus the Maki parameter\cite{maki}, defined by $\alpha_{\mathrm{M}}=\sqrt{2}H_{\mathrm{c2}}^{\mathrm{orb}}(0)/H_\mathrm{P}$, is above 4, suggesting either dominant Pauli-limiting effect at low temperatures or a novel spin-triplet superconductivity. Consistently, measurements using the single crystals\cite{canfield1} reveal two initial slopes of $-$7 and $-$12 T/K for $\mathbf{H}\parallel c$ and $\mathbf{H}\perp c$, respectively. Zuo et al.\cite{zzw} obtained a similar with the slopes of $-$4.2 and $-$10.3 T/K. The data given by Wang et al.\cite{jp} are $-$5 and $-$16.1 T/K, respectively. Therefore, the anisotropy in $H_{\mathrm{c2}}^{\mathrm{orb}}(0)$, defined by $\gamma_{H}^{\mathrm{orb}}(0)=H_{\mathrm{c2}}^{\parallel,\mathrm{orb}}(0)/H_{\mathrm{c2}}^{\bot,\mathrm{orb}}(0)$, is 1.7-3.2, which can be understood in terms of effective-mass anisotropy ($m_{\perp}/m_{\parallel}$ = 3-10). Note that the upper bound of the effective-mass anisotropy is consistent with the resistivity anisotropy value at $T\rightarrow T_\mathrm{c}$.

Balakirev et al.\cite{canfield2} measured $H_{\mathrm{c2}}^{\bot}(T)$ and $H_{\mathrm{c2}}^{\parallel}(T)$ for K$_{2}$Cr$_{3}$As$_{3}$ under magnetic fields up to 60 T, using a contactless technique based on a proximity detector oscillator (PDO). The results show a strong negative curvature for $H_{\mathrm{c2}}^{\parallel}(T)$, characteristic of Pauli-limited behavior. In contrast, $H_{\mathrm{c2}}^{\bot}(T)$ is basically linear, without the Pauli-paramagnetic effect. As a result, $H_{\mathrm{c2}}^{\bot}(T)$ and $H_{\mathrm{c2}}^{\parallel}(T)$ cross at $\sim$4 K, leading to an apparent reversal in the anisotropy, $\gamma_{H}=H_{\mathrm{c2}}^{\parallel}(T)/H_{\mathrm{c2}}^{\bot}(T)$, with decreasing temperature. It is argued that the Pauli-limited behavior of $H_{\mathrm{c2}}^{\parallel}(T)$ is inconsistent with triplet superconductivity. As for the absence of Pauli-limiting effect for $H_{\mathrm{c2}}^{\parallel}(T)$, a form of singlet superconductivity with the electron spins locked along the $c$ direction is proposed\cite{canfield2}.

\begin{figure}
\centering
\includegraphics[width=8cm]{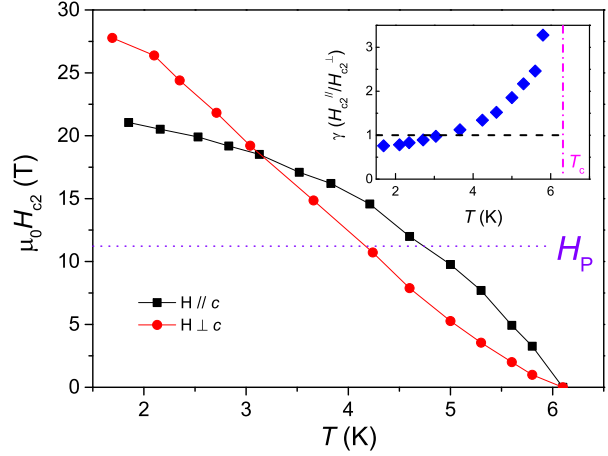}
\caption{Temperature dependence of upper critical fields of K$_{2}$Cr$_{3}$As$_{3}$ crystals for $\mathbf{H}\parallel c$ and $\mathbf{H}\perp c$. $H_\mathrm{P}$ denotes the Pauli-paramagnetic limit. The inset shows temperature dependence of the anisotropy in upper critical fields. Data taken from Ref.\cite{zzw}}
\label{fig8}%\ref{fig:example1}
\end{figure}

Zuo et al.\cite{zzw} report the field-angle and temperature dependence of $H_{\mathrm{c2}}$ for K$_{2}$Cr$_{3}$As$_{3}$ crystals by direct magnetoresistance measurement. Fig.~\ref{fig8} shows the $H_{\mathrm{c2}}^{\bot}(T)$ and $H_{\mathrm{c2}}^{\parallel}(T)$ data, which confirm the result measured by PDO technique. Note that $\gamma_{H}(T)$ tends to diverge when approaching $T_\mathrm{c}$. The reduced $\gamma_{H}(T)$ with decreasing temperature is then due to the Pauli-limiting effect exclusively for $\mathbf{H}\parallel c$. The polar angle $\theta$ dependence of $H_{\mathrm{c2}}$ explicitly indicates absence of Pauli-limiting effect for $\mathbf{H}\perp c$. Besides, the $H_{\mathrm{c2}}$ shows a unique azimuthal angle $\phi$ dependence (a three-fold modulation), which suggests that the Cooper-pair spins cannot be locked along the $c$ axis. The $H_{\mathrm{c2}}$ result of K$_{2}$Cr$_{3}$As$_{3}$ is reminiscent of the anisotropic Pauli-limiting behavior in UPt$_3$\cite{UPt3-Hc2}. The latter is interpreted as an evidence of (psuedo)spin-triplet superconductivity\cite{sauls}. We believe that similar spin-triplet scenario is also the case for K$_{2}$Cr$_{3}$As$_{3}$ and perhaps for Rb$_{2}$Cr$_{3}$As$_{3}$ which shows a similar $\gamma_{H}$ reversal\cite{jp}. It is of interest to see whether there is still an apparent $\gamma_{H}$ reversal in Cs$_{2}$Cr$_{3}$As$_{3}$.

\subsection{Impurity effect}

One of the main arguments opposing triplet superconductivity in K$_2$Cr$_3$As$_3$ is the preliminary observation of insensitivity of $T_\mathrm{c}$ on nonmagnetic impurity scattering as reflected by the residual resistance ratio (RRR) of samples including the polycrystals\cite{canfield1,canfield2}. From the anisotropic resistivity demonstrated above, however, the RRR value of K$_2$Cr$_3$As$_3$ polycrystals is mainly determined by the $\rho^{\perp}(T)$ behavior which has a low RRR value primarily because of the 3D-to-1D crossover at 50 K. Therefore, one needs to clarify the nonmagnetic impurity scattering effect on $T_\mathrm{c}$ with using K$_2$Cr$_3$As$_3$ crystals.

\begin{figure}
\centering
\includegraphics[width=14cm]{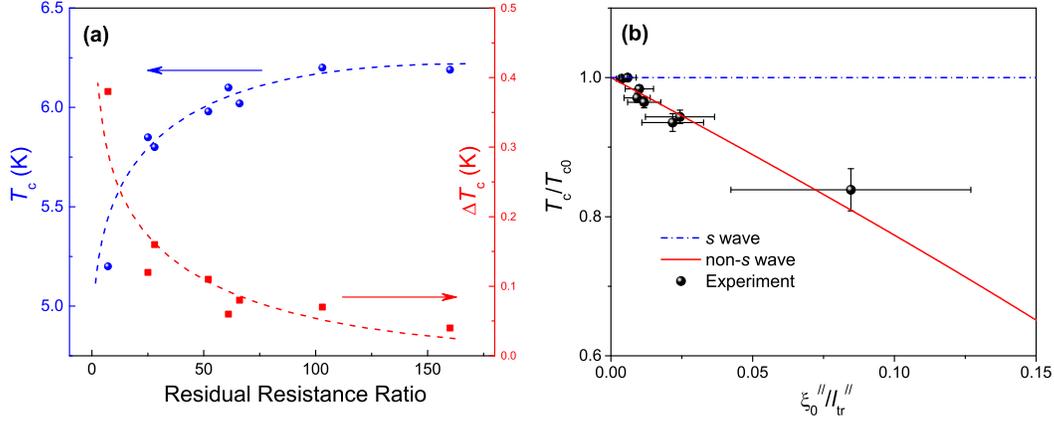}
\caption{(a) Superconducting transition temperature (left axis) and the transition width (right axis) as functions of the residual resistance ratio for K$_{2}$Cr$_{3}$As$_{3}$ crystals. (b)  $T_{\mathrm{c}}/T_{\mathrm{c0}}$ versus $\xi_{0}^{\parallel}/l_{\mathrm{tr}}^{\parallel}$ in K$_{2}$Cr$_{3}$As$_{3}$. See the text for further information. Adapted from Ref.\cite{ly}.}
\label{fig9}%\ref{fig:example1}
\end{figure}

We were able to grow K$_2$Cr$_3$As$_3$ single crystals with remarkably different RRR values, using different purity of Cr as the source material,\cite{ly}. The impurity atoms are Fe, Al, Ga, V, etc., which serve as nonmagnetic scattering centers, because magnetic impurities would induce a resistivity minimum in $\rho(T)$ which is practically absent. Fig.~\ref{fig9}(a) show the $T_{\mathrm{c}}$ and the superconducting transition width, $\Delta T_{\mathrm{c}}$, defined by the temperature difference between 10\% and 90\% values in the extrapolated normal-state resistance. One sees that $T_{\mathrm{c}}$ saturates to the highest value of 6.2 K for large RRRs. When the RRR is below $\sim$25, $T_{\mathrm{c}}$ is reduced rapidly, accompanied with an obvious broadening in the superconducting transition. Since the residual resistivity $\rho_{0}$ is inversely proportional to RRR (the room-temperature resistivity almost keeps constant), as a result, $T_{\mathrm{c}}$ decreases almost linearly with the increase of $\rho_{0}$\cite{ly}. Similar phenomenon is observed in the $p$-wave spin-triplet superconductor Sr$_2$RuO$_4$\cite{Ru}.

According to Anderson's theorem\cite{anderson}, $T_{\mathrm{c}}$ hardly changes by nonmagnetic scattering for an $s$-wave superconductor. Therefore, the above result suggests non-$s$ wave superconductivity in K$_{2}$Cr$_{3}$As$_{3}$. The nonmagnetic scattering effect for non-$s$-wave superconductors takes a similar form of Abrikosov-Gor'kov (AG) pair-breaking equation\cite{AG,Ru},

\begin{equation}
\mathrm{ln}\left(\frac{T_{\mathrm{c0}}}{T_{\mathrm{c}}}\right)=\psi\left(\frac{1}{2}+g\frac{T_{\mathrm{c0}}}{T_{\mathrm{c}}}\right)-\psi\left(\frac{1}{2}\right),
\end{equation}

where $\psi$ is the digamma function, $g=\hbar/(4\pi \tau k_{\mathrm{B}}T_{\mathrm{c0}})$ is a measure of the pair breaking. Here $\tau$ denotes the mean free time due to impurity scattering, which is related to the electron mean free path and Fermi velocity by $\tau=\ell/v_{\mathrm{F}}$. $T_{\mathrm{c0}}$ is the $T_{\mathrm{c}}$ value in the clean limit. Since superconducting coherence length $\xi_{0}$ is proportional to $v_\mathrm{F}/\Delta_{0}$, where $\Delta_{0}\sim T_{\mathrm{c0}}$ is the superconducting gap, hence the $g$ parameter can be quantified by $\xi_{0}/\ell$. The $\xi_{0}^{\parallel}$ value can be obtained by $H_{\mathrm{c2,\parallel}}^{\mathrm{orb}}(0)$, which is about 3.5 nm\cite{zzw}. The electron mean free path along $c$ axis can be estimated by Drude model, $\ell_{\mathrm{tr}}^{\parallel}=129.6\gamma_{\mathrm{N}}/(Z\rho_{0}^{\parallel})$\cite{ly}. Fig.~\ref{fig9}(b) shows the reduced superconducting transition temperature, $T_{\mathrm{c}}/T_{\mathrm{c0}}$, as a function of $\xi_{0}^{\parallel}/l_{\mathrm{tr}}^{\parallel}$. Indeed, the experimental data points basically meet the AG formula, which quantitatively verifies a non-$s$ wave superconductivity in K$_{2}$Cr$_{3}$As$_{3}$.

\subsection{Other physical properties}

Apart from the physical properties described above, $A_2$Cr$_3$As$_3$ show many other peculiar properties by the measurement of specific heat\cite{K233,Rb233,canfield1} and penetration depth\cite{yhq1,yhq2}, nuclear quadrapole resonance (NQR)\cite{imai1,imai2,zgq}, nuclear magnetic resonance (NMR)\cite{imai2,zgq}, and muon-spin relaxation or rotation ($\mu$SR)\cite{adroja1,adroja2}. Below we briefly summarize and remark on these progresses.

\textbf{Specific heat.} The specific-heat measurements on K$_2$Cr$_3$As$_3$ to date\cite{K233,canfield1} give following information. (1) The normal-state electronic specific-heat coefficient (or Sommerfeld coefficient), $\gamma_\mathrm{N}$ = 70-75 mJ K$^{-2}$ mol-fu$^{-1}$, is over three times of the expected one from the `bare' density of states by the first-principles calculations\cite{cc1}, which can be understood in terms of either electron-electron interactions or electron-phonon coupling. (2) The dimensionless specific-heat jump at $T_{\text{c}}$, [$\Delta C/(\gamma T_{\text{c}})$], is as high as 2.4, suggesting a strong coupling scenario. (3) In the superconducting state, there is an upturn for $C_\text{e}(T)/T$ below $\sim$1.0 K, probably due to a Schottky anomaly from the related nuclei and/or impurities. This makes it difficult to study the intrinsic $C_\text{e}(T)$ behavior from which information on the superconducting pairing symmetry can be obtained.

An alternative method that partially removes the influence from the Schottky anomaly is to study the $C(H,T)/T$ data well above the Schottky-anomaly temperature\cite{Rb233}. One can extrapolate the field-induced Sommerfeld coefficient of $\gamma(H) \approx \frac{H}{H_{\mathrm{c2}}(0)}\gamma_{\mathrm{N}}$ at zero temperature in the
superconducting mixed state. We find that $\gamma(H)$ basically follows the Volovik
relation\cite{volovik}, suggesting the presence of nodes in the superconducting gap in Rb$_2$Cr$_3$As$_3$\cite{Rb233}.

\textbf{NQR, NMR, $\mu$SR, etc.}. Imai and co-workers\cite{imai1} report the first $^{75}$As nuclear quadrapole resonance (NQR) study for K$_2$Cr$_3$As$_3$. A strong enhancement of Cr-spin fluctuations above $T_{\mathrm{c}}$ is evidenced based on the nuclear spin-lattice
relaxation rate 1/$T_1$, suggesting non-phonon-mediated superconductivity. They also find a power-law temperature dependence of 1/$T_{1}T\sim T^{-\delta}$ with $\delta\sim$0.25, consistent with the scenario of Tomonaga-Luttinger liquid. In addition, no Hebel-Slichter coherence peak is observed just below $T_\mathrm{c}$, supporting unconventional superconductivity as well.

Zheng and co-workers\cite{zgq} studied the $^{75}$As nuclear magnetic resonance (NMR) as well as $^{75}$As NQR for Rb$_2$Cr$_3$As$_3$. Similarly, they observed enhancement of 1/$T_{1}T$ for $T_{\mathrm{c}}<T<$ 100 K. Simultaneously, the $^{75}$As Knight shift also increases with decreasing temperature, suggesting that the spin fluctuation is of FM. This result is consistent with the theoretical calculations and analysis\cite{cc1,hjp1}, favoring spin-triplet Cooper pairing. In the superconducting state, there is no Hebel-Slichter peak either. The 1/$T_{1}T$ data below 3 K  follows a $T^5$ dependence, suggesting presence of point nodes in the gap function.

The FM spin fluctuations in the normal state revealed by NMR is to some extent consistent with the zero-field $\mu$SR measurements\cite{adroja1}, which show an evidence of the spontaneous appearance of an internal magnetic field below $T_{\mathrm{c}}$. The transverse-field measurement data are better explained for the superconducting gap with line nodes, which agrees with the penetration depth measurement using a tunnel diode oscillator technique\cite{yhq1,yhq2}. The $\mu$SR measurements for Cs$_2$Cr$_3$As$_3$ give a similar result. Nevertheless, NMR investigations on Cs$_2$Cr$_3$As$_3$\cite{imai2} shows lack of enhancement of spin-lattice relaxation rate near $T_\mathrm{c}$. The strength of Cr spin fluctuations, reflected by 1/$T_{1}T$, decrease with increasing the size of alkali-metal ions. This trend seems to correlate with the $\gamma_\mathrm{N}$ and $T_\mathrm{c}$ values [Fig.~\ref{fig4}(b)] altogether\cite{imai2}.

\section{Concluding remarks}\label{sec:5}

As the unique ambient-pressure Cr-based superconductors with Q1D crystal structure, $A_{2}$Cr$_{3}$As$_{3}$ have attracted considerable research interest since their discovery. The main progresses made and the remaining problems are listed as follows.

(1) The crystal structure is basically established. It is characterized by the infinite [(Cr$_3$As$_3$)$^{2-}$]$_{\infty}$ linear chains or DSTs, the center of which is a twist tube made of face-sharing Cr$_6$ octahedra. Noted here is that there are two inequivalent Cr sites in the [(Cr$_3$As$_3$)$^{2-}$]$_{\infty}$ chains, which is in contrast with the unique Cr site in the non-superconducting ``cousins" $A$Cr$_{3}$As$_{3}$. As such, the crystal structure of Rb$_2$Cr$_3$As$_3$ and Cs$_2$Cr$_3$As$_3$ should be re-investigated to verify the relatively length of Cr$-$Cr bond distances.

(2) $T_\mathrm{c}$ is found to decrease monotonically with increasing the interchain distance. Similar trend is also seen for the Sommerfeld coefficient $\gamma_{\mathrm{N}}$ and the NMR $1/T_{1}T$ value near $T_\mathrm{c}$. However, $N(E_\mathrm{F})$ from DFT calculations does not follow this trend. Besides, pressure effect on $T_\mathrm{c}$ even shows the opposite trend on $T_\mathrm{c}$. Variations of these quantities against interchain coupling should be well interpreted in the future.

(3) The resistivity of K$_{2}$Cr$_{3}$As$_{3}$ shows a strikingly anisotropic behavior.  The possible 3D-to-1D dimension crossover (with increasing temperature) is of great interest. Also it is intriguing to verify whether or not the high-temperature state exhibits Luttinger liquid behavior. Besides, anisotropic magnetoresistivity and Hall effect is called for to clarify the nature of the normal state.

(4) Owing to the short superconducting coherence length, K$_{2}$Cr$_{3}$As$_{3}$ samples are normally in the clean limit, hence the $T_{\mathrm{c}}$ value looks robust against impurity scattering. However, for K$_{2}$Cr$_{3}$As$_{3}$ crystals with sufficiently large residual resistivity, the $T_{\mathrm{c}}$ value is reduced remarkably. The $T_{\mathrm{c}}$ depression quantitatively obeys the generalized AG theory, strongly supporting a non-$s$-wave superconductivity. Further investigations on the impurity-scattering effect in Rb$_{2}$Cr$_{3}$As$_{3}$ and Cs$_{2}$Cr$_{3}$As$_{3}$ are useful to make a complete understanding on this issue.

(5) The $\gamma_{\mathrm{N}}$ value for K$_2$Cr$_3$As$_3$ is 70-75 mJ K$^{-2}$ mol-fu$^{-1}$, which is about 4 times of the value from the first-principles calculations. The origin of the mass renormalization, \emph{i.e.}, whether it is due to electron correlations or electron-phonon coupling, needs to be clarified. Also, the lowered $\gamma_{\mathrm{N}}$ for Rb$_{2}$Cr$_{3}$As$_{3}$ and Cs$_{2}$Cr$_{3}$As$_{3}$ needs an explanation.

(4) K$_2$Cr$_3$As$_3$ shows a fully anisotropic Pauli-limiting behavior in the upper critical field $H_{\mathrm{c2}}$, resembling the case in UPt$_3$. The $\mu_{0}H_{\mathrm{c2}}^{\perp}$ value achieves 37 T at 0.6 K, which is over three times larger than the Pauli-paramagnetic limit. Additionally, $H_{\mathrm{c2}}(\phi)$ shows three-fold modulations at low temperatures. While there is no much discrepancy in the experimental result, debates exist on the interpretation\cite{canfield2,zzw}. The disagreements need to be clarified.

(6) The $^{75}$As NQR shows strong enhancement of Cr-spin (FM) fluctuations above $T_{\mathrm{c}}$ and, there is no Hebel-Slichter coherence peak in the temperature dependence of nuclear spin-lattice relaxation rate just below $T_{\mathrm{c}}$ for K$_2$Cr$_3$As$_3$\cite{imai1}. Similar result is given by the NMR for Rb$_2$Cr$_3$As$_3$, from which \emph{ferromagnetic} spin fluctuations are additionally evidenced\cite{zgq}, supporting a spin-triplet Cooper pairing. The latter seems to be consistent with the observation of a spontaneous internal magnetic field near $T_{\mathrm{c}}$, although being very weak, in the muon spin relaxation or rotation ($\mu$SR) experiment\cite{adroja1}. In the future, the Knight shift of the single crystals needs to be measured through $T_\mathrm{c}$.

(7) So far, existence of nodes (either line nodes or point nodes) in the superconducting gap function is demonstrated or suggested by several independent experiments including London penetration depth\cite{yhq1,yhq2}, specific heat\cite{Rb233}, NQR\cite{imai1,zgq}, etc. More techniques such as STM are expected to make it clearer.

(8) Band-structure calculations show that Cr-3$d$ orbitals dominate the electronic states at the Fermi level ($E_\mathrm{F}$) and, the consequent Fermi surface consists of a 3D FS in addition to two Q1D FS in K$_2$Cr$_3$As$_3$\cite{cc1,hjp1,alemany}. Ferromagnetic spin fluctuations are suggested by the calculations\cite{cc1}. Theoretical models\cite{zy,hjp2,djh} are established based on the molecular orbitals, from which spin-triplet superconductivity is mostly stabilized. Future experimental studies using ARPES are highly desirable to directly reveal the band structures and hopefully the pairing symmetry.

As shown in the review paper by Hirsch, Maple and Marsiglio\cite{maple}, USCs are much rare compared with the conventional electron-phonon superconductors. In this paper we see that unconventional superconductivity is very likely for the newly discovered $A_2$Cr$_3$As$_3$ system which hosts both Q1D crystal structure and significant electron correlations. Furthermore, the pairing symmetry could be of a new type. This conference paper calls for more extensive and in-depth works that could answer the open questions in this emerging research topic.

\section*{Acknowledgement}
We thank Z. A. Xu, J. Y. Liu, H. Q. Yuan, Chao Cao, F. L. Ning, T. Imai, Guo-qing Zheng, L. L. Sun, and Z. W. Zhu for fruitful collaborations. Thanks are also due to D. F. Agterberg, F. C. Zhang, Y. Zhou, J. H. Dai, F. Yang, and J. P. Hu for helpful discussions. This work was supported by the National Natural Science Foundation of China (No. 11674281), National Key R \& D Program of the MOST of China (Grant No. 2016YFA0300202), and the Fundamental Research Funds for the Central Universities of China.

\bibliographystyle{tPHM}
\bibliography{233}

\end{document}